\documentclass[twocolumn,tighten]{aastex62}

\usepackage{graphicx}
\usepackage{ulem}
\usepackage{amsmath}
\usepackage{wrapfig}

\usepackage{tikz}
\usepackage{ulem}

\definecolor{green}{rgb}{0.0, 0.5, 0.0}
\definecolor{brightube}{rgb}{0.82, 0.62, 0.91}

\def\Msun{{M_\odot}}

\newcommand\lsim{\mathrel{\rlap{\lower4pt\hbox{\hskip1pt$\sim$}}
        \raise1pt\hbox{$<$}}}
\newcommand\gsim{\mathrel{\rlap{\lower4pt\hbox{\hskip1pt$\sim$}}
        \raise1pt\hbox{$>$}}}

\begin{document}
\shorttitle{Growth of Massive Stars in Star Clusters}
\shortauthors{Gonz\'{a}lez Prieto et al.}

\title{IMBH Progenitors from Stellar Collisions in Dense Star Clusters}

\correspondingauthor{Elena Gonz\'{a}lez Prieto}
\email{elena.prieto@northwestern.edu}

\author[0000-0002-0933-6438]{Elena Gonz\'{a}lez Prieto}
\affil{Center for Interdisciplinary Exploration \& Research in Astrophysics (CIERA) and Department of Physics \& Astronomy, Northwestern University, Evanston, IL 60208, USA}

\author[0000-0002-9660-9085]{Newlin C.~Weatherford}
\affil{Center for Interdisciplinary Exploration \& Research in Astrophysics (CIERA) and Department of Physics \& Astronomy, Northwestern University, Evanston, IL 60208, USA}

\author[0000-0002-7330-027X]{Giacomo Fragione}
\affil{Center for Interdisciplinary Exploration \& Research in Astrophysics (CIERA) and Department of Physics \& Astronomy, Northwestern University, Evanston, IL 60208, USA}

\author[0000-0002-4086-3180]{Kyle Kremer}
\affiliation{TAPIR, California Institute of Technology, Pasadena, CA 91125, USA}

\author[0000-0002-7132-418X]{Frederic A.~Rasio}
\affil{Center for Interdisciplinary Exploration \& Research in Astrophysics (CIERA) and Department of Physics \& Astronomy, Northwestern University, Evanston, IL 60208, USA}

\begin{abstract}

Very massive stars (VMSs) formed via a sequence of stellar collisions in dense star clusters have been proposed as the progenitors of massive black hole seeds. VMSs could indeed collapse to form intermediate-mass black holes (IMBHs), which would then grow by accretion to become the supermassive black holes observed at the centers of galaxies and powering high-redshift quasars. Previous studies have investigated how different cluster initial conditions affect the formation of a VMS, including mass segregation, stellar collisions, and binaries, among others. In this study, we investigate the growth of VMSs with a new grid of Cluster Monte Carlo (\texttt{CMC}) star cluster simulations---the most expansive to date. The simulations span a wide range of initial conditions, varying the number of stars, cluster density, stellar initial mass function (IMF), and primordial binary fraction. We find a gradual shift in the mass of the most massive collision product across the parameter space; in particular, denser clusters born with top-heavy IMFs provide strong collisional regimes that form VMSs with masses easily exceeding $1000\,\Msun$. Our results are used to derive a fitting formula that can predict the typical mass of a VMS formed as a function of the star cluster properties. Additionally, we study the stochasticity of this process and derive a statistical distribution for the mass of the VMS formed in one of our models, recomputing the model 50 times with different initial random seeds. 

\vspace{1cm}
\end{abstract}
\vspace{2cm}

\section{Introduction}
\label{sec:intro}

Although the dynamical evolution of dense star clusters has been studied extensively for decades, the details of the cluster formation stages and their initial conditions remain highly uncertain. Efforts to tackle these open questions are taking place both theoretically and observationally. In particular, the latest cosmological simulations are approaching the resolution necessary to robustly track formation of bound clusters in a range of galaxy types and redshifts \citep[e.g.,][]{Ma2020, Grudic2023, Rodriguez2023}. Even so, resolving cluster formation in these large cosmological simulations remains challenging because of their multi-scale nature. On the observational side, the James Webb Space Telescope (JWST) has opened a new window into star cluster formation, with several studies reporting observations of candidate proto-globular clusters at high-redshifts \citep[e.g.,][]{Vanzella2022, Mowla2022}.

Globular clusters (GCs), being some of the densest environments in the Universe, host numerous exotic objects and transient phenomena arising from strong dynamical interactions, including direct physical collisions. Previous studies have shown that young star clusters, the likely progenitors of GCs, may produce stars with masses greatly exceeding the maximum mass in the stellar initial mass function (IMF) through successive stellar collisions and mergers \citep[e.g.,][]{Sanders1970,Quinlan1990, Lee1987,Ebisuzaki2001,Portegies_Zwart2002}. These so-called very massive stars (VMSs) have been the focus of considerable previous theoretical work because they are natural progenitors for intermediate-mass black holes (IMBHs). 

The collisional process to form a VMS begins immediately after the segregation of the most massive stars deep into the core of the cluster. Due to the Spitzer instability \citep{Vishniac1978}, these stars cannot achieve energy equipartition and the core develops a high velocity dispersion, which promotes stellar collisions. For sufficiently dense clusters, growth will start as a result of stellar collisions and mergers. As the mass of a merger product grows, so does its collision cross-section, resulting in even more collisions. This results in a positive feedback loop that can rapidly produce a VMS of hundreds to thousands of solar masses.

To study VMS formation in star clusters, \cite{Gurkan2004} investigated the effects of mass segregation and core collapse. This was accomplished through the implementation of Monte Carlo simulations in systems where parameters such as the cluster density profile, stellar IMF, and initial star count were systematically varied. This study found that the mass of the collapsing core was always close to ${\sim}10^{-3}$ times that of the total cluster mass. Remarkably, this follows the observed correlation between central BH mass and total host mass in many astrophysical environments \citep{Ferrarese2000}. Note, however, that \cite{Gurkan2004} did not account for the effects of stellar evolution. 

\cite{Freitag2006a,Freitag2006b} performed the first cluster simulations that included precise treatment of stellar collisions and followed the evolution of the cluster and formation of a collisional runaway. Particularly, \cite{Freitag2006b} studied runaway collisions in young star clusters by varying physical parameters such as cluster mass, size, and initial concentration. These studies confirmed that when the core collapse timescale is shorter than the main-sequence (MS) evolution timescale for the most massive stars ($t\approx 3\,{\rm Myr}$), the cluster will undergo a collisional runaway. However, these studies did not incorporate the role of binaries in the runaway process, which have an important role in the evolution of the cores of star clusters.

VMSs are often assumed to be progenitors of IMBHs \citep[e.g.,][]{Ebisuzaki2001,Portegies2004, Gurkan2006, Giersz2015, Mapelli2016}. In an early study, \cite{Ebisuzaki2001} introduced the collisional runaway formation scenario for IMBHs and discussed the possibility that these IMBHs will eventually sink to the Galactic center and be the seeds for super-massive BHs (SMBHs). The possibility that massive collision products could avoid the pair-instability regime and directly collapse into a massive BH \citep[e.g.,][]{Spera2019,DiCarlo2019,DiCarlo2020} has recently been confirmed via hydrodynamic simulations of stellar collisions and the product's ensuing evolution \citep{Costa2022,Ballone2023}. The possible presence of IMBHs at the centers of GCs has also been studied for many years \citep[see][for a review]{Greene2020}. Tentative evidence of massive BHs at the cores of GCs includes velocity dispersion signatures in nearby GCs \citep[e.g.,][]{Noyola2010, Jalali2012, Baumgardt2017}, accretion signatures from radio observations \citep[e.g.,][]{Maccarone2004, Paduano2024}, hypervelocity stars \citep[e.g.,][]{Edelmann2005, Gualandris2007}, observations of ultra-luminous X-ray sources \citep[e.g.,][]{Colbert1999, Farrell2009}, and pulsar acceleration measurements \citep{Kiziltan2017}.

The likelihood of a cluster forming a VMS depends on various physical properties, among which is the IMF, which remains poorly constrained to this day, especially at high stellar masses. Although many studies of GCs assume a canonical Kroupa IMF \citep{Kroupa2001}, observations suggest that it may not be universal \citep[e.g.,][]{DeMarchi2007, Bartko2010, Haghi2017, Wirth2022}. Furthermore, several studies have shown that the IMF strongly impacts the dynamical evolution and survival of GCs \citep[e.g.,][]{Chernoff1990, Chatterjee2017, Giersz2019, Weatherford2021}. 

In particular, \cite{Weatherford2021} explored the impact of the slope of the IMF (at the high-mass end) on the compact object population. This study found that in addition to producing more BHs, clusters with a top-heavy IMF also produce substantially more binary BH (BBH) mergers, especially those involving (or resulting in) production of upper-mass-gap BHs \citep[e.g.,][]{Spera17, Takahashi2018, Marchant2019, Farmer2019} and IMBHs. The latter is due to three factors; top-heavy IMFs produce heavier stars and therefore heavier BHs, but also lead to several times more stellar collisions---due to scaling of stellar radii and gravitational focusing with mass---and more hierarchical mergers.

Another physical parameter that influences the rate of dynamical interactions in star clusters is the primordial binary fraction \citep[e.g.,][]{Heggie2003, Chatterjee2010, Fregeau2007}. Since binaries have a larger interaction cross-section than single stars, they offer a larger effective area for encounters to take place. As shown in previous studies by \cite{Gonzalez2021} and \cite{GonzalezPrieto2022}, increasing the binary fraction for high-mass stars ($M > 15 \Msun$) to $100$\%, more in line with observed binary fractions in the Galactic field \citep[e.g.,][]{Sana2012, Moe2017}, dramatically increases the number of massive stellar collisions and thus results in more massive BHs.

In this paper, we re-examine the formation of VMSs while self-consistently modeling stellar and binary evolution. We cover systematically the parameter space, extending boundaries in cluster size, density, and mass. Furthermore, we fully explore the stochasticity of this process and derive statistical distributions for the masses of the VMSs. In Section~\ref{sec:methods}, we describe the physical prescriptions and parameters varied in this study. In Section~\ref{sec:VMS}, we analyze the formation of VMSs in our models, while Section~\ref{sec:fitting_formulae} presents a simple equation to estimate the most massive star formed through collisions in a cluster. We discuss the resulting BH population in Section~\ref{sec:BH} and present a statistical study of our models in Section~\ref{sec:CMC}. Finally, in Section~\ref{sec:DC}, we discuss the implications and caveats of this study. 

\newpage
\section{Methods}
\label{sec:methods}

\subsection{Cluster Simulations}
We perform our simulations using \texttt{Cluster Monte Carlo} (\texttt{CMC}), a H\'{e}non-type Monte Carlo code that models the evolution of star clusters \citep[see][for the most recent overview]{Rodriguez_2022}. \texttt{CMC} incorporates prescriptions for various physical processes such as two-body relaxation \citep{Joshi2000}, treatment for stellar collisions \citep{Fregeau2007}, and direct integration of small $N$-body strong encounters using \texttt{Fewbody} \citep{Fregeau2004}. Finally, the population synthesis code \texttt{COSMIC} is fully integrated into \texttt{CMC} to treat stellar and binary evolution \citep{Breivik19}.

We run a set of $324$ simulations (listed fully in Table~\hyperref[table:Table1]{\ref{table:Table1}}) that systematically investigate a broad spectrum of initial cluster properties. First, the grid varies the initial number of objects in the cluster---both singles and binaries---in the range $N$\footnote{It is worth noting that as we reach the upper limit of the range for the initial number of objects, the computational time becomes quite expensive, posing a practical challenge to detailed resolution of the high-$N$ parameter space. For instance, one simulation of a cluster with $N=32,r_v=0.5, \alpha_3 = 1.6$ and $f_{b{\rm ,high}} = 1.0$ took $\sim 570$~hours using $52$~CPUs from Northwestern's supercomputer.} $=(4, 8, 16, 32) \times 10^5$. Second, to examine the impact of cluster density, we vary the cluster's initial virial radius $r_v/{\rm pc} = (0.5, 1, 2)$. Both the values for $N$ and $r_v$ are motivated by earlier work demonstrating that clusters with these initial conditions evolve into GCs similar to those observed in the Milky Way \citep{Kremer2020,Rui2021}. 

While past studies have explored the role of the IMF and binary fraction independently, the present work examines their combined effect on the cluster. We assume a typical primordial binary fraction of $f_b=0.05$ for stars born less massive than $15\,\Msun$ and vary the high-mass binary fraction $f_{b{\rm ,high}}=(0.05,0.25,1.0)$ for stars born more massive than $15\,\Msun$. We sample primary stellar masses from the \citet{Kroupa2001} multi-component power-law IMF,
\begin{equation}
\xi(m)\propto
\begin{cases}
{m}^{-1.3}& \text{$0.08\le m/\mathrm{M}_\odot\leq 0.5$}\\
{m}^{-2.3}& \text{$0.5\le m/\mathrm{M}_\odot\leq 1.0$}\\
{m}^{-\alpha_3}& \text{$1.0\le m/\mathrm{M}_\odot\leq 150.0$}\,.
\end{cases}
\label{eq:IMF}
\end{equation}
To vary the IMF, we choose three different values for $\alpha_3=(1.6,2.3,3.0)$, corresponding to the approximate 95\% confidence interval around the canonical value, $\alpha_3=2.3$ \citep{Kroupa2001}. For each set of initial conditions, we run three statistically-independent realizations of the same cluster. See Section~\ref{sec:CMC} for a detailed discussion of the number of realizations necessary to resolve key behavior. 

\vspace{1cm}
\subsection{Physical Prescriptions}
\label{sec:physical_prescriptions}
\textit{Modified Radii Prescriptions for Massive Stars}: Stellar evolution is an active area of research, with many uncertainties, especially in the high-mass regime. \cite{Agrawal2020} studied the uncertainties in massive stellar evolution models by comparing different stellar evolution codes, finding a notable disparity (see their Figure 8) between the current extrapolation of maximum radius for massive stars in the \texttt{Single Stellar Evolution} code (\texttt{SSE}) and \texttt{Modules for Experiments in Stellar Astrophysics} (\texttt{MESA}). In particular, for stars with a zero-age main-sequence (ZAMS) ${>}40\,\Msun$, the predicted maximum stellar radius in \texttt{SSE} (used in \texttt{COSMIC}) is an order of magnitude higher than the one predicted by more detailed stellar evolution models such as \texttt{MESA}.

To correct for this likely overestimation of the stellar radius, we have truncated the radius of any star with a ZAMS mass $M \geq 40\,\Msun$ to a maximum value of $10^3 R_{\odot}$. So if a star in our simulations reaches a stage in its evolution where it is assigned a stellar radius ${>}10^3\,R_{\odot}$, we simply re-scale the radius---and the radii of the core and convective envelope, proportionately---to this prescribed limit. While more precise extrapolations of stellar radii are currently under investigation, this provisional change prevents an artificially large collision cross-section, thereby ensuring a more accurate collision rate. We have rigorously tested these new prescriptions on thousands of stars, confirming that it does not alter their stellar evolution from default \texttt{SSE} assumptions. 

\textit{Stellar Collision Products}: The properties of a collision product depend on the details of the collision and internal structure of the stars. Due to the large uncertainties in isolated high-mass stellar evolution---let alone the hydrodynamic complexities of post-collision evolution---we adopt the conservative choice of setting the total mass of the collision product $M_3$ equal to the sum of the masses of the colliding stars ($M_3 = M_1 + M_2$) for collisions involving two MS stars. This assumption of mass conservation is motivated by hydrodynamic simulations of stellar collisions  in globular cluster-like environments \citep[e.g.,][]{Lombardi1996, Sills2001, Costa2022, Ballone2023}. In the case of a collision between a giant star and a MS star, we assume that the resulting object has the core of the giant star ($M_{c3}$) embedded in the envelope of both stars ($M_3 = M_1 + M_2$ and $M_{c3} = M_{c1}$). 

The product of all stellar collisions must be rejuvenated since new gas is introduced into the envelope and potentially the core of the new star---giving opportunity to burn more fuel. We assign the rejuvenated effective age of the merger product to be
\begin{equation}\label{eq:rejuvenation}
t_{3} = f_{ \rm rejuv} \frac{t_{\rm MS3}}{M_3} \Big(\frac{M_{1}t_{1}}{t_{\rm MS1}} + \frac{M_{2}t_{2}}{t_{\rm MS2}} \Big),
\end{equation}
where $(t_{\rm MS1}, \, t_{\rm MS2}, \, t_{\rm MS3})$ are the MS lifetimes of the two collision components and the collision product while $(t_1, \, t_2)$ are the stellar ages of the two collision components. $f_{\rm rejuv}$ is a coefficient that determines the level of rejuvenation experienced by the collision product. We adopt $f_{\rm rejuv}=1$ by default and refer the reader to \cite{Breivik19} for a discussion of these rejuvenation prescriptions as well as the choice for $f_{\rm rejuv}$.

\startlongtable
\begin{deluxetable*}{l| c c c c c | c c c | c | c}
\tabletypesize{\scriptsize}
\tablewidth{0pt}
\setlength{\tabcolsep}{1.2\tabcolsep}  
\tablecaption{List of cluster models \label{table:Table1}}
\tablehead{
    \colhead{$\rm ^1 Model$} &
	\colhead{$\rm ^2 N$} &
	\colhead{$\rm ^3 r_{v}$} &
	\colhead{$\rm ^4 \alpha_{3}$} &
	\colhead{$\rm ^5 f_{fb, high}$} &
 	\colhead{$ \rm ^6 M_{clus} $} &
	\colhead{$ \rm ^7 M_{\star, max 1} $} &
    \colhead{$ \rm ^8 M_{\star, max 2} $} &
    \colhead{$ \rm ^9 M_{\star, max 3} $} &
    \colhead{$ \rm ^{10} M_{model, max} $} &
	\colhead{$\rm  ^{11} N_{coll} \times f_{massive}$} \\
	\colhead{} &
	\colhead{[$\times 10^{5}]$} &
	\colhead{$\rm [pc]$} &
	\colhead{} &
	\colhead{} &
    \colhead{$ \rm [10^{5} \, \Msun]$} &
	\colhead{$ \rm [\Msun]$} &
    \colhead{$ \rm [\Msun]$} &
    \colhead{$ \rm [\Msun]$} &
    \colhead{$ \rm [\Msun]$} &
	\colhead{$ (\rm M > 15 \Msun)$} \\
    }
\startdata
a0 & 4 & 0.5 & 1.6 & 0.05 & 8.5 & 1059 & 929 & 1502 & $ 530^{+301}_{-210} $ & 5.06 \\ 
a1 & 4 & 0.5 & 1.6 & 0.25 & 9.3 & 1145 & 1348 & 1424 & $ 687^{+302}_{-221} $ & 8.83 \\ 
a2 & 4 & 0.5 & 1.6 & 1.0 & 12.4 & 1691 & 1579 & 1688 & $ 855^{+375}_{-286} $ & 34.97 \\ 
a3 & 4 & 0.5 & 2.3 & 0.05 & 2.4 & 358 & 465 & 306 & $ 300^{+178}_{-113} $ & 0.06 \\ 
a4 & 4 & 0.5 & 2.3 & 0.25 & 2.5 & 475 & 376 & 362 & $ 387^{+184}_{-130} $ & 0.26 \\ 
a5 & 4 & 0.5 & 2.3 & 1.0 & 2.7 & 556 & 661 & 659 & $ 483^{+253}_{-177} $ & 0.89 \\ 
a6 & 4 & 0.5 & 3.0 & 0.05 & 1.6 & 98 & $ 145^{\dagger} $ & $ 130^{\dagger} $ & $ 196^{+138}_{-81} $ & $<0.01$ \\ 
a7 & 4 & 0.5 & 3.0 & 0.25 & 1.6 & 137 & $ 145^{\dagger} $ & 173 & $ 251^{+159}_{-98} $ & 0.01 \\ 
a8 & 4 & 0.5 & 3.0 & 1.0 & 1.7 & 245 & 243 & 219 & $ 313^{+219}_{-132} $ & 0.02 \\ 
b0 & 4 & 1 & 1.6 & 0.05 & 8.5 & $ 149^{\dagger} $ & 156 & 179 & $ 291^{+127}_{-99} $ & 0.18 \\ 
b1 & 4 & 1 & 1.6 & 0.25 & 9.3 & 263 & 336 & 241 & $ 376^{+93}_{-92} $ & 1.43 \\ 
b2 & 4 & 1 & 1.6 & 1.0 & 12.4 & 355 & 295 & 397 & $ 468^{+140}_{-130} $ & 7.47 \\ 
b3 & 4 & 1 & 2.3 & 0.05 & 2.4 & $ 145^{\dagger} $ & 175 & $ 148^{\dagger} $ & $ 164^{+75}_{-53} $ & $<0.01$ \\ 
b4 & 4 & 1 & 2.3 & 0.25 & 2.5 & $ 145^{\dagger} $ & $ 149^{\dagger} $ & 186 & $ 211^{+70}_{-55} $ & 0.02 \\ 
b5 & 4 & 1 & 2.3 & 1.0 & 2.7 & 198 & 290 & 196 & $ 264^{+103}_{-76} $ & 0.09 \\ 
b6 & 4 & 1 & 3.0 & 0.05 & 1.6 & $ 95^{\dagger} $ & $ 145^{\dagger} $ & $ 130^{\dagger} $ & $ 106^{+63}_{-40} $ & $<0.01$ \\ 
b7 & 4 & 1 & 3.0 & 0.25 & 1.6 & $ 95^{\dagger} $ & $ 145^{\dagger} $ & $ 130^{\dagger} $ & $ 139^{+67}_{-48} $ & $<0.01$ \\ 
b8 & 4 & 1 & 3.0 & 1.0 & 1.7 & $ 95^{\dagger} $ & $ 145^{\dagger} $ & $ 130^{\dagger} $ & $ 172^{+98}_{-63} $ & $<0.01$ \\ 
c0 & 4 & 2 & 1.6 & 0.05 & 8.5 & 205 & $ 149^{\dagger} $ & 233 & $ 159^{+87}_{-59} $ & 0.04 \\ 
c1 & 4 & 2 & 1.6 & 0.25 & 9.3 & 245 & 190 & 252 & $ 206^{+85}_{-62} $ & 0.18 \\ 
c2 & 4 & 2 & 1.6 & 1.0 & 12.4 & 281 & 250 & 268 & $ 257^{+113}_{-83} $ & 1.33 \\ 
c3 & 4 & 2 & 2.3 & 0.05 & 2.4 & $ 145^{\dagger} $ & $ 149^{\dagger} $ & $ 148^{\dagger} $ & $ 89^{+50}_{-33} $ & $<0.01$ \\ 
c4 & 4 & 2 & 2.3 & 0.25 & 2.5 & $ 145^{\dagger} $ & $ 149^{\dagger} $ & 220 & $ 116^{+53}_{-37} $ & $<0.01$ \\ 
c5 & 4 & 2 & 2.3 & 1.0 & 2.7 & 232 & 243 & 199 & $ 145^{+76}_{-50} $ & 0.01 \\ 
c6 & 4 & 2 & 3.0 & 0.05 & 1.6 & $ 95^{\dagger} $ & $ 145^{\dagger} $ & $ 130^{\dagger} $ & $ 59^{+39}_{-25} $ & $<0.01$ \\ 
c7 & 4 & 2 & 3.0 & 0.25 & 1.6 & $ 95^{\dagger} $ & $ 145^{\dagger} $ & $ 130^{\dagger} $ & $ 76^{+46}_{-30} $ & $<0.01$ \\ 
c8 & 4 & 2 & 3.0 & 1.0 & 1.7 & $ 95^{\dagger} $ & $ 145^{\dagger} $ & $ 130^{\dagger} $ & $ 95^{+66}_{-40} $ & $<0.01$ \\ 
\hline 
d0 & 8 & 0.5 & 1.6 & 0.05 & 17.1 & 1947 & 1194 & 549 & $ 637^{+365}_{-236} $ & 8.83 \\ 
d1 & 8 & 0.5 & 1.6 & 0.25 & 18.7 & 1460 & 1806 & 1718 & $ 820^{+342}_{-253} $ & 19.32 \\ 
d2 & 8 & 0.5 & 1.6 & 1.0 & 24.8 & 2310 & 1950 & 2167 & $ 1027^{+490}_{-337} $ & 84.59 \\ 
d3 & 8 & 0.5 & 2.3 & 0.05 & 4.8 & 482 & 656 & 504 & $ 360^{+188}_{-126} $ & 0.14 \\ 
d4 & 8 & 0.5 & 2.3 & 0.25 & 5.0 & 728 & 420 & 817 & $ 465^{+190}_{-136} $ & 0.42 \\ 
d5 & 8 & 0.5 & 2.3 & 1.0 & 5.4 & 502 & 526 & 549 & $ 582^{+273}_{-187} $ & 1.73 \\ 
d6 & 8 & 0.5 & 3.0 & 0.05 & 3.3 & 152 & 200 & 170 & $ 237^{+145}_{-90} $ & $<0.01$ \\ 
d7 & 8 & 0.5 & 3.0 & 0.25 & 3.3 & 159 & 190 & 201 & $ 304^{+161}_{-102} $ & 0.01 \\ 
d8 & 8 & 0.5 & 3.0 & 1.0 & 3.3 & 203 & 334 & 235 & $ 379^{+225}_{-145} $ & 0.05 \\ 
e0 & 8 & 1 & 1.6 & 0.05 & 17.1 & 230 & 186 & 217 & $ 350^{+147}_{-110} $ & 0.61 \\ 
e1 & 8 & 1 & 1.6 & 0.25 & 18.7 & 345 & 322 & 252 & $ 451^{+119}_{-100} $ & 2.27 \\ 
e2 & 8 & 1 & 1.6 & 1.0 & 24.8 & 358 & 327 & 331 & $ 566^{+189}_{-143} $ & 16.39 \\ 
e3 & 8 & 1 & 2.3 & 0.05 & 4.8 & $ 149^{\dagger} $ & $ 148^{\dagger} $ & $ 149^{\dagger} $ & $ 196^{+72}_{-56} $ & 0.01 \\ 
e4 & 8 & 1 & 2.3 & 0.25 & 5.0 & 217 & 350 & $ 149^{\dagger} $ & $ 254^{+62}_{-49} $ & 0.04 \\ 
e5 & 8 & 1 & 2.3 & 1.0 & 5.4 & 230 & 231 & 265 & $ 317^{+112}_{-78} $ & 0.19 \\ 
e6 & 8 & 1 & 3.0 & 0.05 & 3.3 & $ 134^{\dagger} $ & $ 126^{\dagger} $ & $ 137^{\dagger} $ & $ 128^{+62}_{-42} $ & $<0.01$ \\ 
e7 & 8 & 1 & 3.0 & 0.25 & 3.3 & $ 134^{\dagger} $ & $ 126^{\dagger} $ & $ 137^{\dagger} $ & $ 166^{+62}_{-47} $ & $<0.01$ \\ 
e8 & 8 & 1 & 3.0 & 1.0 & 3.3 & $ 134^{\dagger} $ & 181 & $ 137^{\dagger} $ & $ 207^{+103}_{-66} $ & $<0.01$ \\ 
f0 & 8 & 2 & 1.6 & 0.05 & 17.1 & $ 149^{\dagger} $ & 267 & 167 & $ 190^{+104}_{-70} $ & 0.12 \\ 
f1 & 8 & 2 & 1.6 & 0.25 & 18.7 & 258 & 251 & 238 & $ 248^{+107}_{-74} $ & 0.28 \\ 
f2 & 8 & 2 & 1.6 & 1.0 & 24.8 & 365 & 311 & 321 & $ 309^{+146}_{-99} $ & 2.57 \\ 
f3 & 8 & 2 & 2.3 & 0.05 & 4.8 & 230 & 201 & $ 149^{\dagger} $ & $ 108^{+51}_{-38} $ & $<0.01$ \\ 
f4 & 8 & 2 & 2.3 & 0.25 & 5.0 & 255 & 252 & $ 149^{\dagger} $ & $ 139^{+55}_{-40} $ & $<0.01$ \\ 
f5 & 8 & 2 & 2.3 & 1.0 & 5.4 & 227 & 256 & 211 & $ 173^{+85}_{-55} $ & 0.01 \\ 
f6 & 8 & 2 & 3.0 & 0.05 & 3.3 & $ 134^{\dagger} $ & 158 & $ 137^{\dagger} $ & $ 70^{+41}_{-27} $ & $<0.01$ \\ 
f7 & 8 & 2 & 3.0 & 0.25 & 3.3 & $ 134^{\dagger} $ & 158 & $ 137^{\dagger} $ & $ 91^{+47}_{-31} $ & $<0.01$ \\ 
f8 & 8 & 2 & 3.0 & 1.0 & 3.3 & 212 & $ 126^{\dagger} $ & $ 137^{\dagger} $ & $ 113^{+68}_{-42} $ & $<0.01$ \\ 
\hline 
g0 & 16 & 0.5 & 1.6 & 0.05 & 34.4 & 657 & 925 & 449 & $ 761^{+447}_{-298} $ & 18.18 \\ 
g1 & 16 & 0.5 & 1.6 & 0.25 & 37.6 & 1763 & 2424 & 2049 & $ 987^{+494}_{-325} $ & 47.94 \\ 
g2 & 16 & 0.5 & 1.6 & 1.0 & 49.9 & 2507 & 2700 & 2874 & $ 1241^{+739}_{-423} $ & 285.44 \\ 
g3 & 16 & 0.5 & 2.3 & 0.05 & 9.7 & 542 & 492 & 655 & $ 427^{+218}_{-150} $ & 0.31 \\ 
g4 & 16 & 0.5 & 2.3 & 0.25 & 9.9 & 647 & 858 & 542 & $ 552^{+225}_{-160} $ & 0.77 \\ 
g5 & 16 & 0.5 & 2.3 & 1.0 & 10.8 & 942 & 791 & 1176 & $ 696^{+351}_{-224} $ & 2.92 \\ 
g6 & 16 & 0.5 & 3.0 & 0.05 & 6.6 & $ 149^{\dagger} $ & 177 & 157 & $ 281^{+162}_{-102} $ & 0.01 \\ 
g7 & 16 & 0.5 & 3.0 & 0.25 & 6.6 & 222 & 387 & 227 & $ 364^{+183}_{-116} $ & 0.02 \\ 
g8 & 16 & 0.5 & 3.0 & 1.0 & 6.7 & 529 & 309 & 335 & $ 454^{+266}_{-165} $ & 0.1 \\ 
h0 & 16 & 1 & 1.6 & 0.05 & 34.4 & 274 & 249 & 216 & $ 420^{+198}_{-137} $ & 1.63 \\ 
h1 & 16 & 1 & 1.6 & 0.25 & 37.6 & 317 & 401 & 383 & $ 541^{+205}_{-143} $ & 6.65 \\ 
h2 & 16 & 1 & 1.6 & 1.0 & 49.9 & 385 & 458 & 483 & $ 673^{+289}_{-197} $ & 43.26 \\ 
h3 & 16 & 1 & 2.3 & 0.05 & 9.7 & $ 149^{\dagger} $ & 161 & $ 149^{\dagger} $ & $ 236^{+85}_{-67} $ & 0.01 \\ 
h4 & 16 & 1 & 2.3 & 0.25 & 9.9 & 195 & 272 & 236 & $ 303^{+77}_{-62} $ & 0.07 \\ 
h5 & 16 & 1 & 2.3 & 1.0 & 10.8 & 352 & 371 & 302 & $ 378^{+141}_{-103} $ & 0.42 \\ 
h6 & 16 & 1 & 3.0 & 0.05 & 6.6 & $ 149^{\dagger} $ & $ 142^{\dagger} $ & $ 136^{\dagger} $ & $ 154^{+67}_{-47} $ & $<0.01$ \\ 
h7 & 16 & 1 & 3.0 & 0.25 & 6.6 & $ 149^{\dagger} $ & $ 142^{\dagger} $ & $ 136^{\dagger} $ & $ 199^{+68}_{-51} $ & $<0.01$ \\ 
h8 & 16 & 1 & 3.0 & 1.0 & 6.7 & $ 149^{\dagger} $ & 155 & 140 & $ 248^{+121}_{-76} $ & 0.01 \\ 
i0 & 16 & 2 & 1.6 & 0.05 & 34.4 & 233 & 180 & 196 & $ 230^{+139}_{-87} $ & 0.3 \\ 
i1 & 16 & 2 & 1.6 & 0.25 & 37.6 & 291 & 244 & 271 & $ 297^{+156}_{-99} $ & 1.19 \\ 
i2 & 16 & 2 & 1.6 & 1.0 & 49.9 & 257 & 333 & 309 & $ 372^{+220}_{-129} $ & 6.16 \\ 
i3 & 16 & 2 & 2.3 & 0.05 & 9.7 & $ 149^{\dagger} $ & 243 & $ 149^{\dagger} $ & $ 129^{+63}_{-44} $ & $<0.01$ \\ 
i4 & 16 & 2 & 2.3 & 0.25 & 9.9 & 241 & 182 & 238 & $ 166^{+73}_{-49} $ & $<0.01$ \\ 
i5 & 16 & 2 & 2.3 & 1.0 & 10.8 & 216 & 263 & 192 & $ 208^{+102}_{-70} $ & 0.03 \\ 
i6 & 16 & 2 & 3.0 & 0.05 & 6.6 & $ 149^{\dagger} $ & $ 142^{\dagger} $ & $ 136^{\dagger} $ & $ 84^{+46}_{-31} $ & $<0.01$ \\ 
i7 & 16 & 2 & 3.0 & 0.25 & 6.6 & $ 149^{\dagger} $ & 179 & $ 136^{\dagger} $ & $ 109^{+55}_{-36} $ & $<0.01$ \\ 
i8 & 16 & 2 & 3.0 & 1.0 & 6.7 & $ 149^{\dagger} $ & 190 & $ 136^{\dagger} $ & $ 136^{+84}_{-50} $ & $<0.01$ \\ 
\hline 
j0 & 32 & 0.5 & 1.6 & 0.05 & 68.6 & 757 & 796 & 695 & $ 920^{+674}_{-387} $ & 47.75 \\ 
j1 & 32 & 0.5 & 1.6 & 0.25 & 75.1 & 3427 & 4251 & 4553 & $ 1184^{+775}_{-447} $ & 131.15 \\ 
j2 & 32 & 0.5 & 1.6 & 1.0 & 99.6 & 5545 & 26864 & 3590 & $ 1479^{+1072}_{-588} $ & 950.73 \\ 
j3 & 32 & 0.5 & 2.3 & 0.05 & 19.3 & 827 & 1057 & 894 & $ 518^{+289}_{-188} $ & 0.55 \\ 
j4 & 32 & 0.5 & 2.3 & 0.25 & 19.8 & 770 & 701 & 887 & $ 663^{+335}_{-210} $ & 1.68 \\ 
j5 & 32 & 0.5 & 2.3 & 1.0 & 21.5 & 1491 & 1083 & 973 & $ 828^{+517}_{-294} $ & 7.08 \\ 
j6 & 32 & 0.5 & 3.0 & 0.05 & 13.1 & 161 & 282 & 220 & $ 340^{+197}_{-124} $ & 0.01 \\ 
j7 & 32 & 0.5 & 3.0 & 0.25 & 13.1 & 344 & 264 & 307 & $ 438^{+227}_{-143} $ & 0.04 \\ 
j8 & 32 & 0.5 & 3.0 & 1.0 & 13.3 & 360 & 466 & 619 & $ 546^{+359}_{-202} $ & 0.19 \\ 
k0 & 32 & 1 & 1.6 & 0.05 & 68.6 & 242 & 347 & 262 & $ 501^{+323}_{-191} $ & 5.79 \\ 
k1 & 32 & 1 & 1.6 & 0.25 & 75.1 & 371 & 286 & 334 & $ 651^{+349}_{-217} $ & 16.91 \\ 
k2 & 32 & 1 & 1.6 & 1.0 & 99.6 & 542 & 508 & 451 & $ 810^{+500}_{-287} $ & 109.65 \\ 
k3 & 32 & 1 & 2.3 & 0.05 & 19.3 & 231 & 213 & 179 & $ 283^{+126}_{-89} $ & 0.04 \\ 
k4 & 32 & 1 & 2.3 & 0.25 & 19.8 & 276 & 268 & 175 & $ 365^{+138}_{-96} $ & 0.17 \\ 
k5 & 32 & 1 & 2.3 & 1.0 & 21.5 & 342 & 288 & 348 & $ 456^{+222}_{-141} $ & 1.04 \\ 
k6 & 32 & 1 & 3.0 & 0.05 & 13.1 & $ 147^{\dagger} $ & $ 145^{\dagger} $ & $ 147^{\dagger} $ & $ 184^{+84}_{-56} $ & $<0.01$ \\ 
k7 & 32 & 1 & 3.0 & 0.25 & 13.1 & $ 147^{\dagger} $ & 166 & $ 147^{\dagger} $ & $ 239^{+92}_{-66} $ & $<0.01$ \\ 
k8 & 32 & 1 & 3.0 & 1.0 & 13.3 & 169 & $ 145^{\dagger} $ & 232 & $ 300^{+161}_{-99} $ & 0.01 \\ 
l0 & 32 & 2 & 1.6 & 0.05 & 68.6 & 305 & 251 & 180 & $ 274^{+205}_{-113} $ & 0.95 \\ 
l1 & 32 & 2 & 1.6 & 0.25 & 75.1 & 235 & 265 & 291 & $ 357^{+238}_{-135} $ & 2.44 \\ 
l2 & 32 & 2 & 1.6 & 1.0 & 99.6 & 320 & 321 & 381 & $ 443^{+319}_{-180} $ & 15.78 \\ 
l3 & 32 & 2 & 2.3 & 0.05 & 19.3 & $ 149^{\dagger} $ & $ 149^{\dagger} $ & 232 & $ 154^{+90}_{-56} $ & $<0.01$ \\ 
l4 & 32 & 2 & 2.3 & 0.25 & 19.8 & 220 & 186 & 180 & $ 201^{+106}_{-68} $ & 0.02 \\ 
l5 & 32 & 2 & 2.3 & 1.0 & 21.5 & 209 & 174 & 195 & $ 248^{+161}_{-91} $ & 0.09 \\ 
l6 & 32 & 2 & 3.0 & 0.05 & 13.1 & $ 147^{\dagger} $ & $ 145^{\dagger} $ & $ 147^{\dagger} $ & $ 101^{+57}_{-37} $ & $<0.01$ \\ 
l7 & 32 & 2 & 3.0 & 0.25 & 13.1 & 165 & $ 145^{\dagger} $ & 219 & $ 131^{+70}_{-46} $ & $<0.01$ \\ 
l8 & 32 & 2 & 3.0 & 1.0 & 13.3 & 210 & $ 145^{\dagger} $ & 220 & $ 163^{+110}_{-63} $ & $<0.01$
\enddata 
\tablecomments{
Columns $2$--$6$ list the initial physical parameters of our clusters, including the initial number of objects, virial radius, absolute value of the high-mass stellar IMF slope ($\alpha_3$), high-mass binary fractio, and cluster mass. Columns $7$--$9$ list the mass of the most massive star formed in each realization of the same model. The dagger indicates masses that result from stellar IMF alone (i.e., those that do not experience any collisional growth). Column $10$ lists the mass of the most massive star as predicted by the fitting formula described in Section~\ref{sec:fitting_formulae}, with the error bars indicating the $95 \%$ confidence interval. Finally, column $11$ records the number of binary--single and binary--binary interactions involving the merger of at least two stars that are both more massive than $15\,\Msun$, normalized by the initial number of massive stars sampled in the cluster. This gives a rough sense of the number of collisions per massive star in each model.}
\end{deluxetable*}

\section{VMS Formation}
\label{sec:VMS}

\begin{figure*}
    \includegraphics[width=0.8\textwidth]{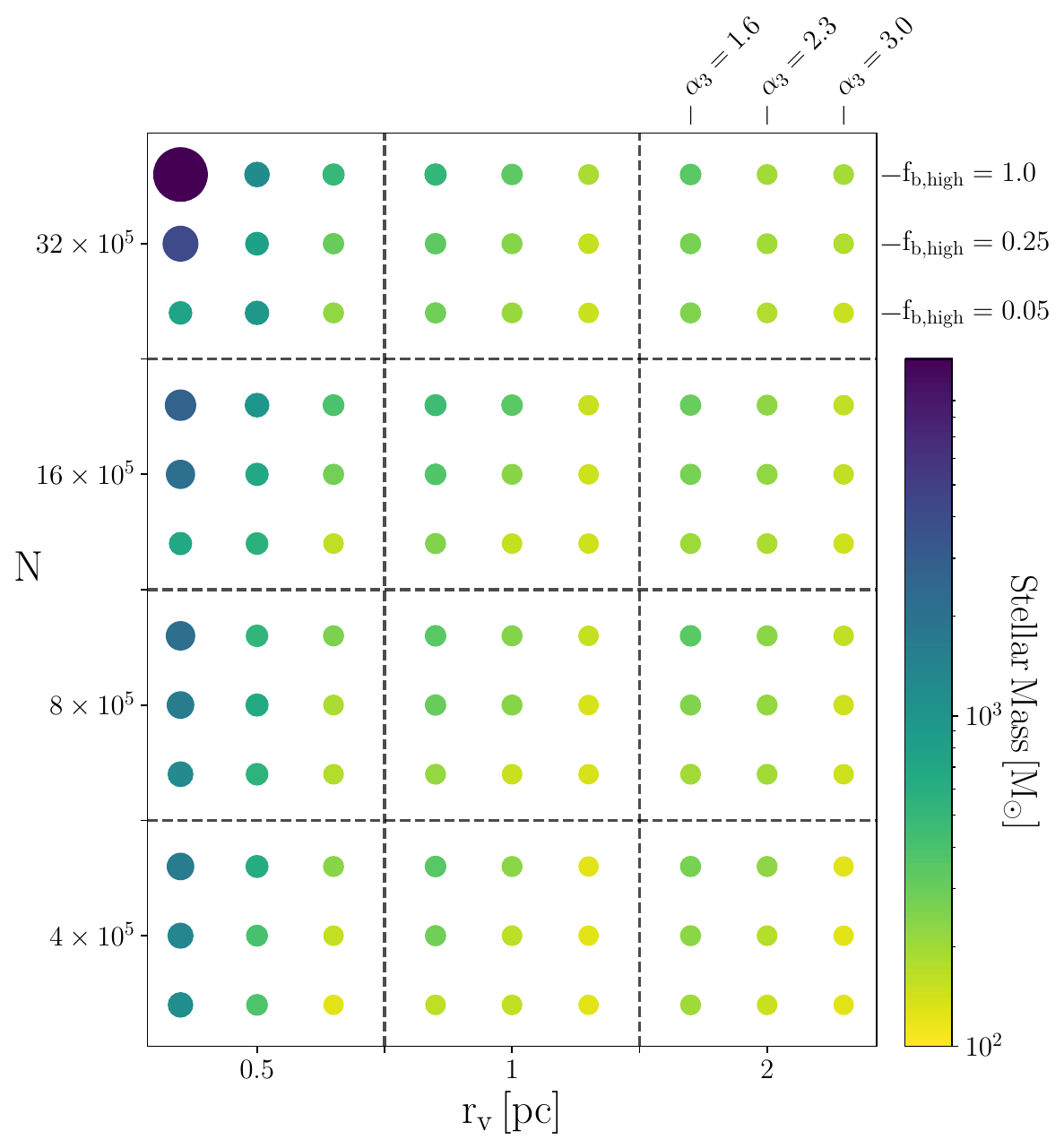}
    \centering
    \caption{\label{fig:max_star} The mean mass of the most massive star formed in a cluster based on its initial conditions. Specifically, the horizontal axis specifies the cluster's initial virial radius $r_v$, and the vertical axis specifies its initial number of objects $N$ (both singles and binaries). Within each box in the $r_v$--$N$ grid, we present a 3x3 sub-grid to distinguish models with different high-mass stellar IMF slope $\alpha_3=(1.6,2.3,3.0)$, from left to right, and high-mass binary fraction $f_{b{\rm ,high}}=(0.05,0.25,1.0)$, from bottom to top. The size and color of the circles reflect the mass of the most massive star formed at each set of initial conditions (averaged over all three realizations).}
\end{figure*}

To study how the formation of a VMS depends on the initial conditions of a star cluster, we closely examine the formation process of the most massive star in each cluster, denoted as $M_{\star, \rm max}$. Figure~\hyperref[fig:max_star]{\ref{fig:max_star}} shows the mean mass of the most massive star formed through stellar collisions across the three realizations performed at each point in the model grid. The plot reveals a consistent trend: as the number of initial objects increases and the cluster becomes more compact (indicated by a smaller $r_v$ value), $M_{\star, \rm max}$ also rises. Furthermore, clusters born with a top-heavy IMF ($\alpha_3=1.6$) feature a collision rate in the first $10$~Myr that is ${\approx}4$ times higher than those born with a canonical IMF (for typical GCs born with $N=8\times 10^5$). As a consequence, for a given combination of $N$ and $r_v$ within the grid, a more top-heavy stellar IMF (smaller $\alpha_3$) results in higher values of $M_{\star, \rm max}$, since a higher number of stellar collisions facilitates the growth of the VMS.

A comparable correlation occurs in the case of the high-mass binary fraction, where a higher primordial fraction of massive binaries increases the mass of the VMS. This is a result of an increased rate of massive star collisions due to the presence of massive binaries in the cluster, which ultimately facilitate the formation of a more massive star. Furthermore, since massive stars are rare compared to low-mass stars, increasing the high-mass binary fraction does not significantly increase the total binary fraction. Consequently, clusters will not experience significant heating from the addition of these primordial massive binaries alone and the process to form the VMS can proceed uninterrupted. The trend is less pronounced for lower-$N$ runs and lower-density clusters, which tend to yield more diffuse clusters where the collisional rate is reduced. Overall, it is evident that the slope of the IMF and the virial radius have the strongest effect on the collisional formation of a VMS in a star cluster.

By examining each section within the parameter space more closely, we can learn more about the physical processes driving the formation of the massive star. Focusing on models with a virial radius of $2$~pc, we find that $M_{\star, \rm max}$ never exceeds $400\,\Msun$ across all initial conditions. The formation of the massive star typically involves a few stellar interactions, or in most cases, one binary--single or binary--binary interaction resulting in the collision of more than two stars. These less-concentrated models, unlike their denser counterparts, do not have as strong of a correlation between $M_{\star, \rm max}$ and the slope of the IMF or the high-mass binary fraction. This can be explained by massive binaries taking a longer time to segregate towards the cluster center, thus limiting their ability to significantly increase the collision rate and trigger a runaway process. 

Models with virial radius of $1$~pc are initially more dense, allowing us to begin observing the onset of the formation of a VMS via a sequence of stellar collisions. The trend becomes apparent as the slope of the IMF becomes shallower, which is equivalent to a leftward movement within each $\rm 3x3$ box in Figure~\hyperref[fig:max_star]{\ref{fig:max_star}}. Most notably, in models with $\alpha_3 = (1.6, 2.3)$ and $f_{b{\rm ,high}} = 1$, the formation of $M_{\star, \rm max}$ primarily occurs through a few collisions (typically between $2$ or $3$) that involve massive stars. In contrast with the formation channel described for the $2$~pc models, the stars involved in these collisions are slightly more massive than those initially sampled from the IMF. These unusually-massive stars tend to be products of binary coalescences, which increase as the  primordial fraction of binaries increases. When combined with a slightly denser cluster, this mechanism facilitates the formation of a VMS.

Finally, models with initial virial radii of $ 0.5$~pc represent the densest clusters in the grid. Within this subset, we observe a greater diversity of pathways leading to the formation of massive stars. In models with a top-light IMF ($\alpha_3 = 3.0$), the progenitors of massive stars experience multiple collisions both during their MS stage and their giant phase. These collisions typically result in a mass gain of approximately $16\,\Msun$ per collision. Most importantly, these collisions tend to occur at later cluster ages, roughly $t>4$~Myr. For models with a canonical IMF, the average mass gain is ${\approx}37\,\Msun$ per collision. A detailed analysis of the collisional histories of $M_{\star, \rm max}$ reveals that the significant (and fast) mass accumulation primarily arises from a series of massive binary-mediated interactions resulting in the merger of more than two stars. This enables a very rapid increase in mass and forms a single VMS on timescales shorter than $4\,{\rm Myr}$. Finally, in models with a top-heavy IMF ($\alpha_3 = 1.6$), we typically observe more than one VMS forming on a timescale $<3$~Myr. In these clusters, multiple massive stars promptly scatter with each other in the center of the cluster and merge, triggering the start of a more extreme runaway process. The average mass gained per collision in these models is ${\approx}71\,\Msun$. 

\begin{figure}
    \centering
    \includegraphics[width=\linewidth]{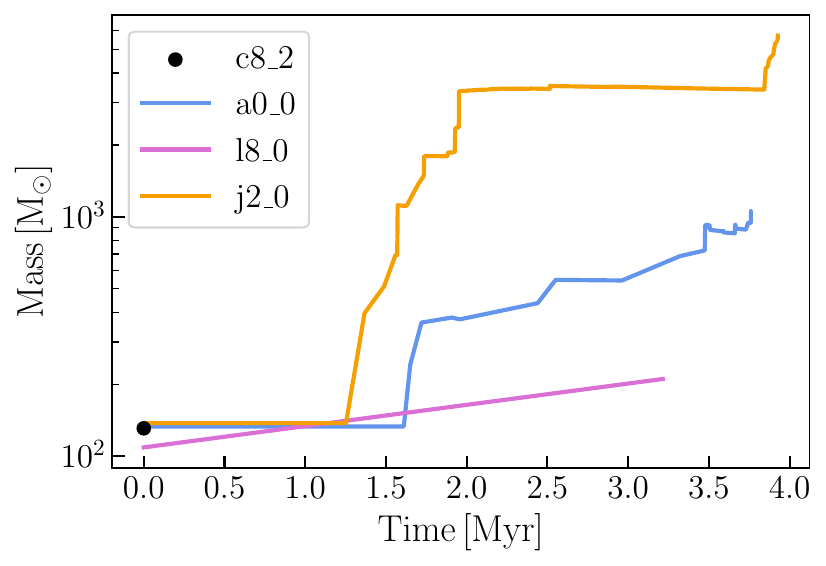}
    \caption{\label{fig:Mass_time}The collision history for the most massive star formed in models $\rm c8, a0, l8$ and $\rm j2$ listed in Table~\hyperref[table:Table1]{\ref{table:Table1}}. The number following the underscore indicates the realization (chosen to be the one that yields the median VMS mass). The massive star formed in model $\rm c8$ (shown as a dot) does not experience any growth.}
\end{figure}

In Figure~\hyperref[fig:Mass_time]{\ref{fig:Mass_time}} we show the collisional history for $M_{\star, \rm max}$ across models spanning from the least dense and least massive to the most dense and most massive (represented by the four corners of Figure~\hyperref[fig:max_star]{\ref{fig:max_star}}). Among each set of three realizations, we selected the one yielding the median VMS mass as a representative case. From the figure, we see that the star that experiences the most growth is not necessarily the most massive star initially sampled from the stellar IMF (shown by the initial mass of the VMS in model $\rm l8$, which is below the IMF's upper bound). Furthermore, the star in model $\rm c8$ (indicated by a dot) does not experience any growth via collisions. The star in model $\rm l8$ also does not experience significant growth, as that model only differs from $\rm c8$ in the initial number of objects. As we move towards denser and initially more top-heavy IMF models shown in yellow and blue, we see the formation of a VMS via successive stellar collisions. Particularly, we begin see the exponential growth of a star in ${<}3$~Myr. 

\begin{figure*}
    \centering
    \includegraphics[width=0.8\linewidth]{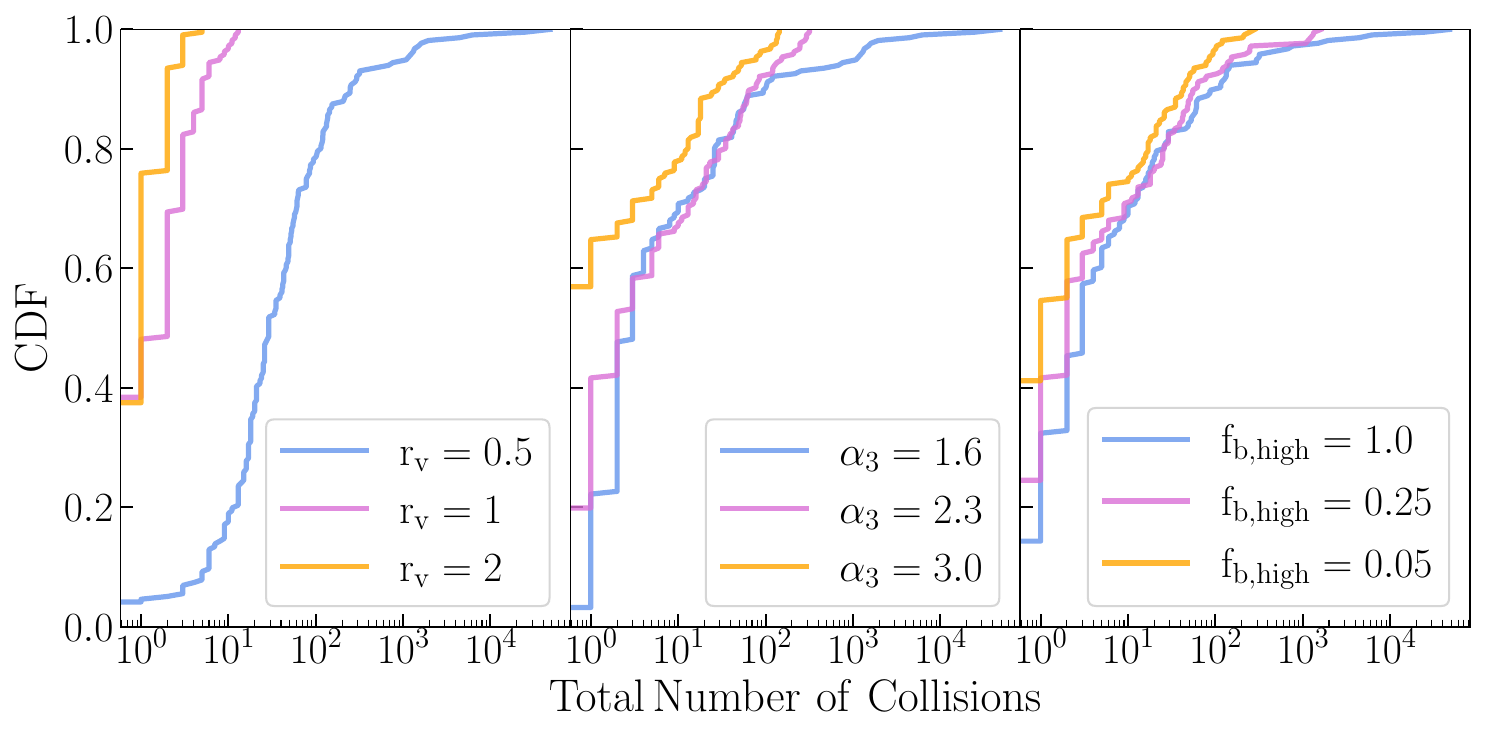}
    \caption{\label{fig:Ncoll}The cumulative distribution of the total number of collisions that contributed to the formation of the most massive star in each of the models. Each panel shows the distribution across all model realizations, differentiated by virial radius (left panel), high-mass stellar IMF slope (center), and high-mass binary fraction (right).}
\end{figure*}

To further illustrate how the channels for forming $M_{\star, \rm max}$ vary across different regions of the parameter space, Figure~\hyperref[fig:Ncoll]{\ref{fig:Ncoll}} depicts the cumulative distribution of the total number of stellar collisions leading to the formation of $M_{\star, \rm max}$ in each model. In the leftmost panel, models with lower $r_v$ consistently exhibit a richer dynamical history, aligning with the expectations that denser clusters will experience higher collision rates. The middle panel reveals a subtler trend, but the overarching message remains that a top-heavy IMF enhances the collision rate. In the rightmost panel, although the trend appears less distinct for varying binary fraction values (as expected from the results shown in  Figure~\hyperref[fig:max_star]{\ref{fig:max_star}}), we still observe that a higher binary fraction enhances the collision rate. It is important to emphasize that we only fix one physical parameter per panel, so the range of total number of stellar collisions exhibited by each line is a consequence of the diverse formation channels across the entire gird.

\begin{figure} 
\centering
\includegraphics[width=1\linewidth]{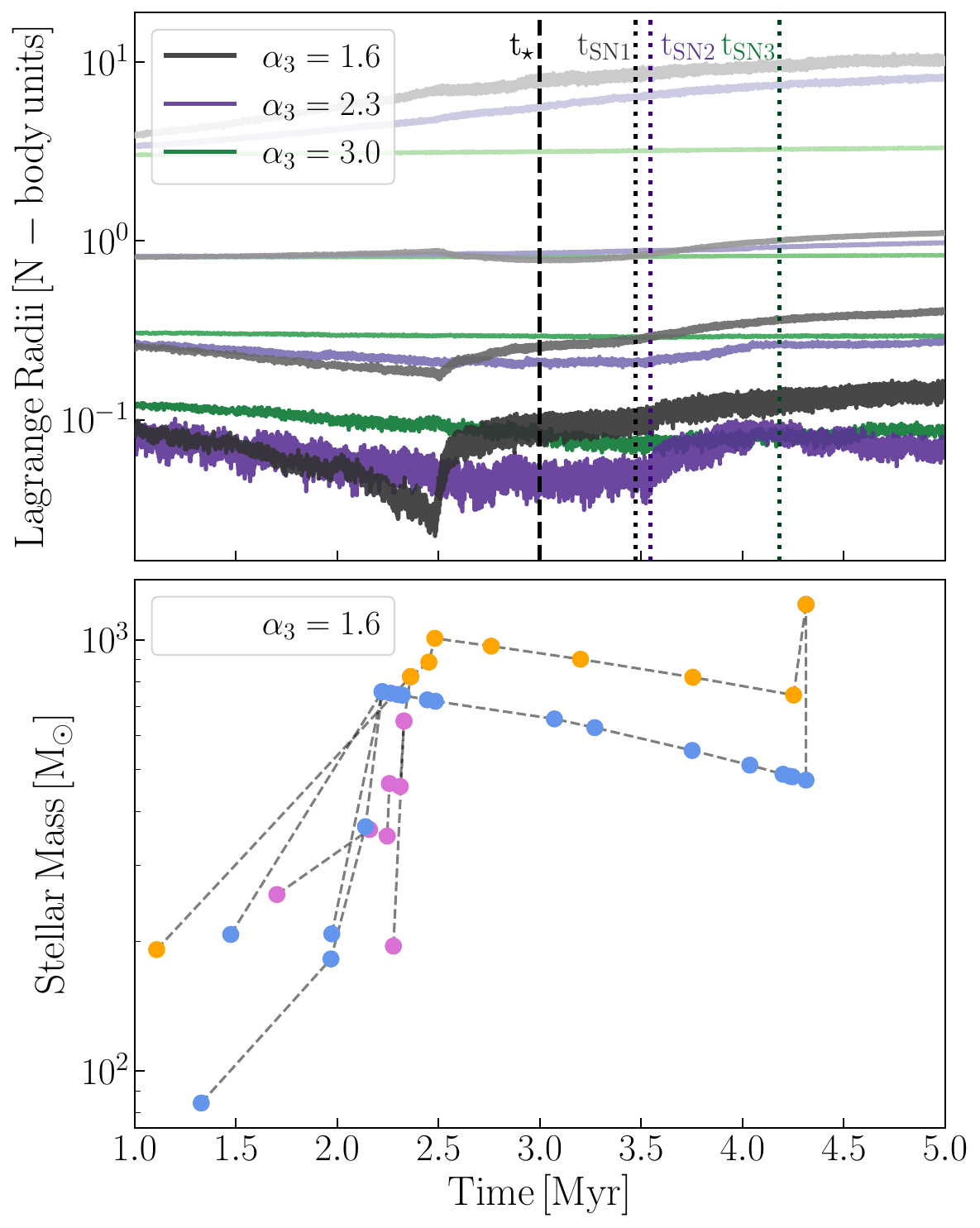}
\caption{\label{fig:Lagrange_radii}\textit{Upper Panel}: The time evolution of the Lagrange radii--- from top to bottom enclosing the $99$\%, $50$\%, $10$\%, and $1$\% of the cluster's total mass---for simulations ($a1,a4,a7$) in Table~\hyperref[table:Table1]{\ref{table:Table1}}. These models all have $N=4\times 10^5$, $r_v=0.5\,{\rm pc}$, and $f_{b{\rm ,high}}=0.25$. We vary $\alpha_3=(1.6,2.3,3.0)$ shown in black, purple, and green curves, respectively. The dashed vertical line represents the main-sequence lifetime for the most massive stars in the cluster ($t=3\,{\rm Myr}$). The dotted vertical lines show the time of the first supernova in each cluster model. \textit{Lower Panel}: The collisional history of the three most massive bodies in the simulation with a top-heavy IMF ($\alpha_3=1.6$), distinguished by color. Two of these massive stars (blue and purple) merge at $t\approx 2.3\,{\rm Myr}$ and the remnant (colored blue) merges with the third massive star (yellow) at $t\approx 4.4\,{\rm Myr}$ to form the final VMS.}
\end{figure}

To gain a more detailed understanding of the dynamics within the core of the clusters that form a VMS, we closely analyze in the top panel of Figure~\hyperref[fig:Lagrange_radii]{\ref{fig:Lagrange_radii}} the evolution of the Lagrange radii (enclosing the specified percentages of the cluster's mass). We pay particular attention to the evolution of objects within the $1$\% Lagrange radius in models $a1$, $a4$, and $a7$, which differ only in $\alpha_3$ (and thereby cluster mass). Notably, the behavior of the innermost particles demonstrates that a top-heavy IMF results in a more pronounced core contraction, which facilitates the formation of a VMS. Although we show only the $\alpha_3$-dependence of the Lagrange radii evolution in Figure~\hyperref[fig:Lagrange_radii]{\ref{fig:Lagrange_radii}} (with a fixed value for $N$, $r_v$, and $f_{b,{\rm high}}$), similar trends exist when comparing models with different $N$, $r_v$, or $f_{b,{\rm high}}$. In particular, a deeper collapse is typically seen for models with smaller $r_v$.

It can also be seen in Figure~\ref{fig:Lagrange_radii} that in models where a very massive star forms, the initial core collapse occurs on a timescale shorter than the MS lifetime for the most massive stars ($t_\star \approx 3\,{\rm Myr}$, marked as a vertical dashed line). This agrees with previous studies that found a runaway process only occurs when $t_{\rm cc} < t_\star$ \cite[e.g.,][]{Freitag2006a}. After $3$~Myr, mass loss due to supernovae and corresponding remnant ejections form natal kicks halts core collapse. This leads to a re-expansion of the core, resulting in a decrease in density, effectively preventing a runaway process. To account for this, we have included the time of the first supernova for each cluster in Figure~\ref{fig:Lagrange_radii} (shown as vertical dotted lines).  Except for the model with $\alpha_3 = 1.6$, the core begins to re-expand as the first supernova event occurs, averting an extreme collisional runaway.

For the model with $\alpha_3 = 1.6$ (and many dense clusters), the initial re-expansion is due to a combination of factors. First, as the core contracts, many objects sink towards the center of the cluster. This increase in central density forms new binaries through three-body binary formation, which in turn heats the core \citep{Cohn1989}. Furthermore, the collisions that formed the massive object have extracted some of the (negative) potential energy of the cluster. Thus, once the series of collisions stops, the cluster begins to re-expand. In cases where a very massive star does not form, like model $a4$ with a canonical stellar IMF (purple curve), it is a combination of mass loss due to supernovae and three-body binary formation that drives the core re-expansion.

To demonstrate how the initial core collapse contributes to the formation of a VMS, we present the collisional history of $M_{\star, \rm max}$ in run $a1$ in the lower panel of Figure~\ref{fig:Lagrange_radii}. Each distinct colored curve within the diagram represents a separate branch in the evolutionary process. At about $2.3$~Myr, three separate massive stars form, each weighing approximately $700\,\Msun$. This occurs precisely when the core undergoes its initial contraction. The stars formed in the yellow and magenta pathways merge, giving rise to a star of roughly $1000\,\Msun$. While the core begins to re-expand, the merger product undergoes gradual mass loss due to stellar winds. At approximately $4.5$~Myr, this star collides with a star of mass $400\,\Msun$, resulting in a collision product with a mass of approximately $1100\,\Msun$.

It is important to note that at any given time in a cluster, more than one VMS might be present, which is not depicted in the figures in this paper. Our study focuses on assessing the extent of runaway phenomena occurring within the initial $10$~Myr of the cluster's lifetime. This takes into account whether or not the massive stars formed in the cluster have enough time to sink to the cluster center due to dynamical friction and merge.

In general, although VMSs form during the initial core collapse (core contraction) in our cluster simulations, the runaway process halts once the core re-expands. As a result, we are not in a regime \citep[unlike previous studies, e.g.,][]{Freitag2006a, Freitag2006b} where an extreme collisional runaway scenario causes the entire cluster core to collapse and form a VMS with a mass ${\sim}10^{-3} $ times the cluster mass. Instead, we observe smaller-scale runaways that enable the cluster to re-expand and continue its evolution. This can be primarily attributed to the delay in the core collapse timescale because of updates in stellar evolution prescriptions and  the role of binaries in heating up the cluster core. Consequently, none of our models reach core collapse during the initial $10$~Myr. The future fate of such clusters (beyond the $10$~Myr modeled in this paper) falls outside the scope of our current study, but will be investigated in a subsequent publication.

\section{Fitting formulae} 
\label{sec:fitting_formulae}

Carefully mapping the different evolutionary outcomes of clusters across a broad physical spectrum requires an extensive, high-resolution grid of simulations. However, this task is rather computationally impractical, so we develop a simple fitting formula that can be used to estimate $M_{\star, \rm max}$ for a cluster, based upon the model grid explored in this paper. We begin with a simple power-law formula for the dependence of the maximum stellar mass on $N$, $r_v$, $\alpha_3$, and $f_{b{\rm ,high}}$:
\begin{equation} \label{eq:nested_sampling}
\frac{M_{\star{\rm ,max}}}{\Msun}
= A \cdot \left(\frac{N}{10^5}\right)^{\eta} \left(\frac{r_v}{\rm pc}\right)^{-\beta} ({\alpha_3})^{-\omega} (f_{b{\rm ,high}})^{\gamma}.
\end{equation}
To determine the values of the coefficient $A$ and the power-law exponents, we perform a Bayesian inference technique known as nested sampling \cite[see][for a review of the method]{Skilling2006}, computing the parameters constraining the model using the \texttt{nestle} package \citep{Shaw2007, Mukherjee2006, Feroz2009}. This method restricts mass priors by sampling within likelihood contours, demonstrating excellent efficacy in high-dimensional parameter estimation. We refer the reader to the Appendix to learn more about the performance of the fit. 

We use a uniform prior from $10^{-3}$--$10^3$ for $A$ and from $10^{-3}$--$10^1$ for each of the exponents. We find that the data is best described by the values in Table~\hyperref[table:Table2]{\ref{table:Table2}}, which shows that the parameters that have the most significant effect on the estimation of $M_{\star, \rm max}$ are $r_v$ and $\alpha_3$. This is expected since a smaller virial radius leads to a higher overall collision rate and a shorter mass segregation timescale for the massive stars, promoting earlier and more frequent massive collisions.

\begin{deluxetable}{c | c | c }
    \tabletypesize{\scriptsize}
    \setlength{\tabcolsep}{0.7\tabcolsep}  
    \centering
    \tablecaption{\label{table:Table2}Best-fit Values for Equation~(\ref{eq:nested_sampling})}  
    \tablehead{
    	\colhead{Parameter} &
    	\colhead{Value}  &
        \colhead{Corresponding Variable} } 
    \startdata
    $\rm A$  &  $716 \pm  184 $ & $ \text{N/A}$ \\  
    $\eta$  &   $0.26  \pm   0.13  $ & $\rm N$ \\  
    $\beta$  &  $ 0.87 \pm   0.23 $ & $\rm r_v$\\  
    $\omega $ &  $ 1.59 \pm   0.39 $ & $\rm \alpha_3$\\  
    $\gamma$  &  $ 0.16  \pm   0.09$ & $f_{b{\rm ,high}}$
    \enddata
    \tablecomments{The best-fit values for the fitting formulae shown in Equation~(\ref{eq:nested_sampling}) obtained using the nested sampling method \citep{Skilling2006}. }
\end{deluxetable}

\begin{figure*}
    \centering
    \includegraphics[width=0.8\textwidth]{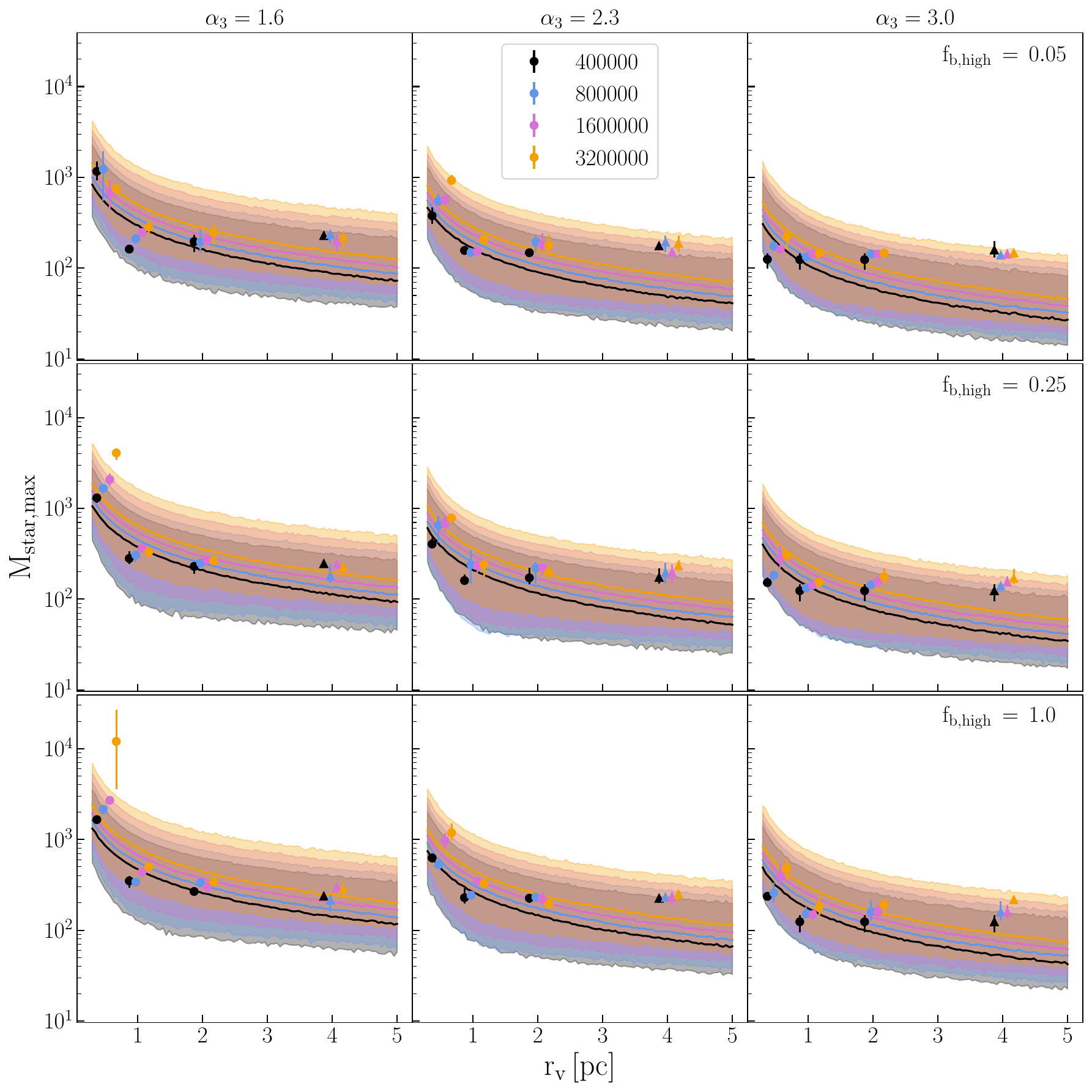}
    \caption{\label{fig:panel_comp} The maximum stellar mass as a function of initial cluster parameters, as given by the fitting formula Equation~(\ref{eq:nested_sampling}). The shaded region represents the $95\%$ credibility region. The mean of the \texttt{CMC} data is over-plotted, with the error bars indicating the maximum and median values obtained in our models. Additional models 
    that were not utilized in the parameter estimation process but are used to demonstrate the performance of the predictive model are shown in triangles.
    Supplementary versions of this Figure with the axes instead showing $\alpha_3$ and $f_{b,{\rm high}}$ are available as a Figure Set in the online journal.}
\end{figure*}

With this model to predict the onset of VMS formation, we aim to extend predictions across a broader range of cluster initial conditions. Figure~\hyperref[fig:panel_comp]{\ref{fig:panel_comp}} shows the predicted maximum stellar mass distribution across models of diverse $f_{b,{\rm high}}$, $r_v$, and $\alpha_{3}$ parameters. To obtain predictive model outcomes, we extract $1000$ samples for every data point across the parameter space by randomly drawing from the parameter distribution listed in Table~\ref{table:Table2}. Subsequently, we calculate the mean of the 1000 samples as well as the $95\%$ credibility region, illustrated in Figure~\hyperref[fig:panel_comp]{\ref{fig:panel_comp}} as a solid line and shaded region, respectively.

To qualitatively demonstrate the performance, we overplot the data obtained from the \texttt{CMC} simulations, revealing strong agreement as the majority of the data falls within the $95$\% confidence interval. Discrepancies between the model and data occur in regions where the initiation of the collisional runaway becomes highly stochastic, leading to substantial variance in the mass of the most massive star. This is the case for the model with $N = 32 \times 10^5$, $r_v = 0.5$~pc, $\alpha_3 = 1.6$, and $f_{b,\rm high} = 1$ (model $\rm j2$ in Table~\hyperref[table:Table1]{\ref{table:Table1}}), where the stochasticity is apparent in the three data points for $M_{\star, \rm max}$.

To assess the accuracy of the predictive model, we computed additional simulations with $ r_v = 4 $~pc, shown in Figure~\hyperref[fig:panel_comp]{\ref{fig:panel_comp}}. These simulations were not utilized in the parameter estimation process for determining the best-fit model. Instead, they are only used to demonstrate the performance of the model. We see that the mean masses from our CMC simulations fall within the $95\%$ confidence interval of the predictive model. We also note that the model under performs in some clusters. This discrepancy arises from the fact that in our simulations, if the cluster does not experience a significant number of collisions, the most massive star at any given time typically corresponds to the most massive star initially sampled from the Kroupa IMF, often around $150 \, \Msun$. Thus, it is to be expected that \texttt{CMC} models with low densities and top-light IMFs will not form collisional products with masses much higher than the IMF upper limit. This is shown by the flat evolution of the maximum stellar mass as a function of virial radius of the \texttt{CMC} data plotted in Figure~\hyperref[fig:panel_comp]{\ref{fig:panel_comp}}. Currently, our fitting formulae does not take into account the assumed IMF, so the prediction for the maximum mass is allowed to go below the IMF upper limit.

It is important to note that this fitting formula is intrinsic to our assumed stellar evolution prescriptions. As such, it may not hold for clusters with different stellar treatments and physical assumptions (e.g., an Elson profile), or for clusters that deviate significantly from the parameter space covered by our model suite (e.g., very low-mass or very high-mass clusters). Moreover, in the denser clusters, the mechanism through which a massive star forms is stochastic, so we expect considerable variance in mass within those regimes.

\section{Massive BH Formation}
\label{sec:BH}

\begin{figure*}
    \centering
    \includegraphics[width=1\textwidth]{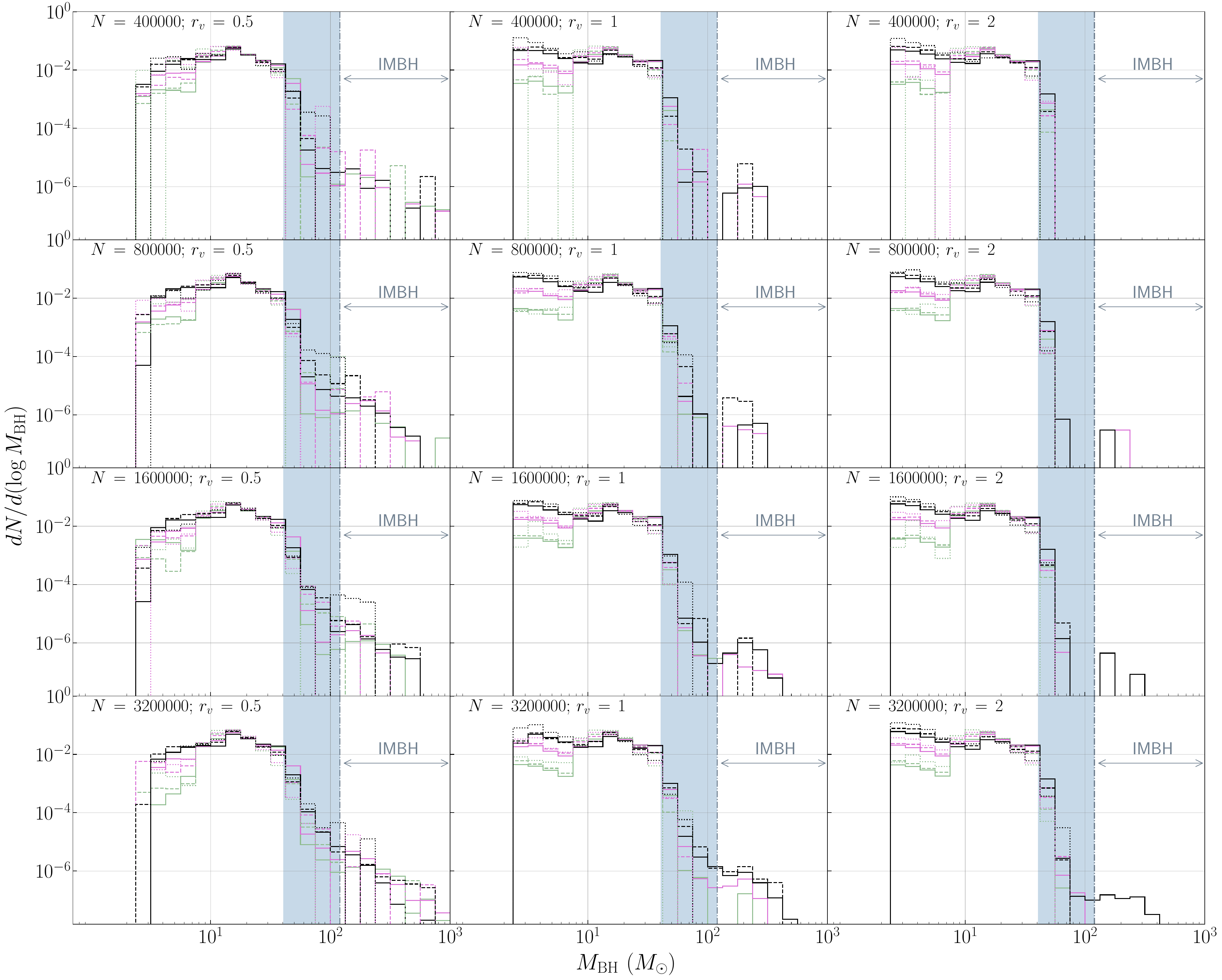}
    \caption{\label{fig:bh_spectrum}Normalized BH mass distribution for all models listed in Table~\hyperref[table:Table1]{\ref{table:Table1}}. Models with $f_{b,{\rm high}} = 1$ are depicted in black, those with $f_{b,{\rm high}} = 0.25$ in magenta, and models with $f_{b,{\rm high}} = 0.05$ in green. The solid lines represent models with $\alpha_3 = 1.6$, dashed lines denote models with  $\alpha_3 = 2.3$, and dotted lines correspond to models with  $\alpha_3 = 3.0$. The blue-shaded region indicates the upper-mass-gap (defined here between $40$--$120\,\Msun$) while an arrow marks the start of the IMBH regime ($M>120\,M_\odot$). }
    
\end{figure*}

An essential question stemming from this research is the fate of the massive stars in these clusters, and whether the formation of the runaway object has an impact on the compact object population.
In Figure~\hyperref[fig:bh_spectrum]{\ref{fig:bh_spectrum}}, we show the spectrum of BH masses formed across the first $10$~Myr of our cluster models. For all values of $\rm N$, as $r_v$ decreases (moving left in each row), there's a notable increase in the number of BHs formed within or beyond the upper--mass-gap \citep[assumed here to be between $40$--$120 \Msun$, but boundaries are uncertain; e.g.,][]{Spera17, Takahashi2018, Farmer2019}. This agrees with the findings of \cite{GonzalezPrieto2022} who showed that denser clusters exhibit higher rates of stellar collisions, facilitating the formation of BHs in the upper-mass-gap and IMBH regimes. 

When increasing the initial number of objects while maintaining a constant $r_v$ (moving downward in each column), we observe that the total number of massive BHs formed increases for dense models. This is due to the increased ``mass budget'' as the number of initial objects grows, which allows more stars that were not previously massive BH progenitors to merge and populate the massive BH region. This is also apparent in the decrease in BHs with masses in the range $4$--$10\,\Msun$. Furthermore, within each panel, a higher binary fraction often correlates with the formation of more massive BHs, alongside a lower $\alpha_{3}$ value. This is also consistent with the general trends we observe in Figure~\hyperref[fig:max_star]{\ref{fig:max_star}}.

Crucially, these BH spectra solely represent the initial $10$~Myr of the cluster's evolution and do not constitute a complete sample of the full BH population. In fact, due to the rejuvenation prescriptions outlined in Section~\ref{sec:methods} and lower-mass stars, there are BH progenitors left in most clusters at that time. We defer more detailed analyses of the long-term population and retention of compact objects to future studies. Nonetheless, the presence of such a diverse population of massive compact objects in these clusters hints at the possibility they could significantly contribute to numerous gravitational-wave events \citep{Rodriguez2019,Kremer20,Gonzalez2021, Weatherford2021, GonzalezPrieto2022}.

\section{CMC Statistical Analysis}
\label{sec:CMC}

\begin{figure}
\centering
\includegraphics[width=1\linewidth]{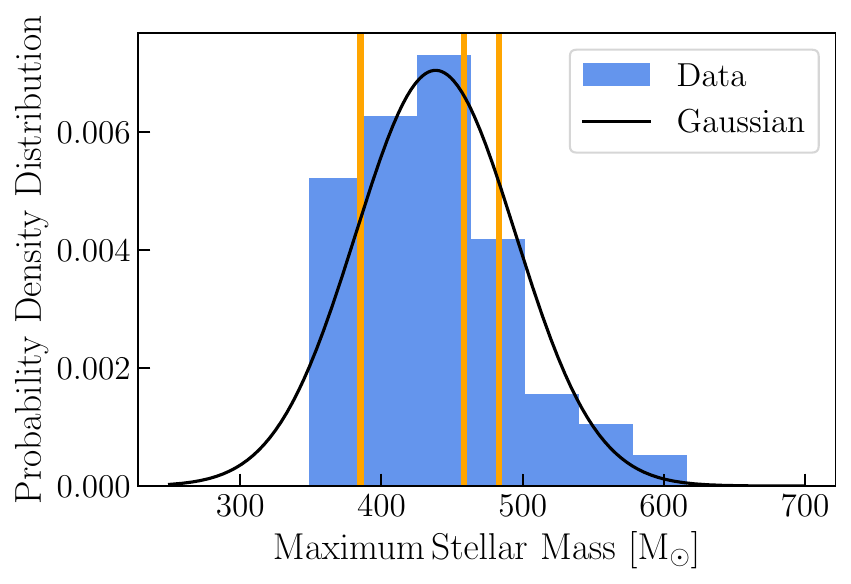}
\caption{\label{fig:dist_masses} Distribution of the maximum VMS mass across all of the $50$ realizations of the model with $N = 1.6 \times 10^6 , \,  r_v = 1 \rm \,  pc, \, \alpha_3 = 1.6$ and $ f_{b{\rm ,high}} = 1.0$. This captures the stochasticity in the outcomes of the collisional runaway. The values from the first three runs are shown in orange. A Gaussian with the same mean and standard deviation is shown in black.}
\end{figure}

Given the inherently statistical nature of the Monte Carlo algorithm, we now investigate whether simulations with identical macroscopic initial conditions but different random seeds (which set the exact initial stellar positions and velocities) yield a statistical mass distribution for $M_{\star, \rm max}$. In Figure~\hyperref[fig:dist_masses]{\ref{fig:dist_masses}}, we present the distribution of stellar masses from a set of $50$ simulations with an initial population of $16 \times 10^5$ objects, a virial radius of $1$~pc, $\alpha_3 = 1.6$, and a high-mass binary fraction of $1$. For qualitative comparison, we also overplot the first three runs in orange. These specific initial conditions were chosen because they represent one of the densest regions in our current grid, often resulting in the formation of a VMS with a mass of a few hundred $\Msun$. While these models are to some extent stochastic in nature, an examination of 
Figure~\hyperref[fig:dist_masses]{\ref{fig:dist_masses}} reveals that the stellar masses roughly follow a Gaussian distribution centered at $440\,\Msun$, with a spread of $50\,\Msun$ (the Gaussian is added for illustrative purposes, using the mean and standard deviation of the data). While the distribution appears to cover a broad range of stellar masses, it is much narrower than the spread of masses shown across the entire grid explored in this study (see Figure~\hyperref[fig:max_star]{\ref{fig:max_star}}).

\begin{figure}
\centering
\includegraphics[width=1\linewidth]{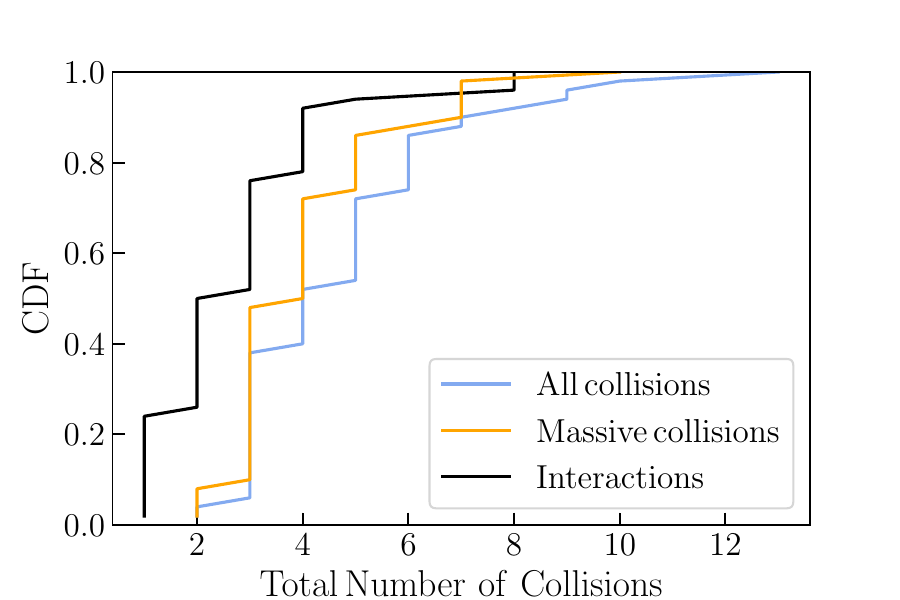}
\caption{\label{fig:Ncoll_dist} The cumulative distribution of the total number of collisions that contributed to the formation of the most massive VMS in each of the $50$ realizations of the model is shown in blue. The orange line specifically denotes the total number of collisions where the colliding star had a mass $ \geq 15\,\Msun$. The total number of interactions the star experienced is shown in black.}
\end{figure}

In all $50$~realizations, the formation of the massive object results from a series of stellar collisions occurring during binary--single and binary--binary interactions. Figure~\hyperref[fig:Ncoll_dist]{\ref{fig:Ncoll_dist}} illustrates the cumulative distribution of the total number of collisions that contributed to the formation of $M_{\star, \rm max}$ in each of the $50$~cluster models. To emphasize the collisions that significantly contribute to mass buildup, we show in orange those collisions where the colliding star is more massive than $15\,\Msun$. As depicted, for most runs, the massive star forms after a few (${\approx}3$--$5$) stellar collisions. In numerous cases, more than one star merges during a binary-mediated interaction. To account for this, we also plot in black the number of interactions that resulted in collisions. It is evident that the number of interactions follows a narrow distribution with a mean of approximately $2.7$ and a standard deviation of $1.68$. Thus, across all $50$ simulations, there is a high level of consistency in the number of interactions that $M_{\star, \rm max}$ undergoes. This marks the first time we have been able to characterize the realization-to-realization variability of \texttt{CMC} models with high resolution, even in the densest and most stochastic regimes. 
The consistent agreement in the low number of collisions required for the formation of a VMS emphasizes the importance of studies using precise hydrodynamic simulations to understand and model the properties of collision products \citep{Costa2022, Ballone2023}.

\vspace{1cm}
\section{Discussion and Conclusions}
\label{sec:DC}
\subsection{Summary}

This paper presents findings from an extensive grid of \texttt{CMC} simulations tracking the dynamical evolution of star clusters during the first $10$~Myr of their lifetime across a spectrum of initial conditions. We particularly focus on the formation process of VMSs, which are likely progenitors of IMBHs. The results from this study can be condensed into three principal findings:

1. Clusters that start with sufficiently high densities experience a phase of core contraction at early times. If this contraction precedes the first supernovae in the cluster, it leads to the formation of a VMS through a collisional runaway instability. In order of importance (see Table~\ref{table:Table2}), the maximum mass reached depends strongly on the high-mass slope of the stellar IMF, the initial cluster density, and the high-mass primordial binary fraction.

2. We have derived a fitting formula that can be used to estimate the mass of the VMS as a function of initial cluster conditions. While this equation depends on specific assumptions regarding stellar evolution and collision prescriptions, it serves as a useful tool to evaluate the potential for a collisional runaway before performing computationally-expensive $N$-body simulations. 

3. At the end of our simulations, some of the VMSs have collapsed to form a BH in the upper-mass-gap or an IMBH. These massive BHs will sink to the center of the cluster and participate in dynamical encounters that will result in interesting signatures such as tidal disruption events, and binary mergers. In particular, BBH mergers containing a more massive component are potentially very important LIGO/LISA sources \citep{Rodriguez2019, Kremer20, Gonzalez2021, Weatherford2021, GonzalezPrieto2022}.

Observations of very high-redshift quasars have sparked debates concerning the formation mechanisms for supermassive BHs. Various proposed channels that can explain the rapid formation and growth of supermassive BHs include the direct collapse of massive Population III stars \citep[e.g.,][]{Stacy2012, Hirano2017, Kimura2021} or massive clouds \citep[e.g.,][]{Loeb1994,Oh&Haiman, Mayer2010}, or the formation of BH seeds via repeated stellar mergers \citep[e.g.,][]{Quinlan1990,Portegies_Zwart2002, Devecchi2009, Tagawa_2020} or hierarchical stellar-mass BH mergers in dense clusters \citep[e.g.,][]{Davies2011, Kroupa2020,Atallah2023}.

The clusters modeled in this work could be similar to the massive star clusters that are believed to be proto-GCs. These clusters are thought to be massive and dense, with low metallicity. Our models can thus help constrain the initial properties of clusters that might be the birth place of the seeds for the very high-redshift quasars observed by many telescopes, including JWST.

\subsection{Caveats and Future Work}

The study of the evolution of single massive stars is currently an active area of research. As an added layer of complexity, most massive stars are observed in close binary systems, making their modelling even more challenging \citep[e.g.,][]{Sana2012a,Moe2017}. As a consequence, we must make assumptions when modeling the evolution of these massive stars. A crucial parameter is their stellar radii, which is poorly constrained for massive stars. As outlined in Section~\hyperref[sec:methods]{\ref{sec:physical_prescriptions}}, this has been partially addressed in this study by the re-scaling of the stellar radii, but more accurate modeling is needed.

Even more uncertain are properties of the collision products. This aspect is particularly relevant to this study, since we investigate the formation of massive stars resulting from numerous stellar collisions. Here, we adopt the simple  ``sticky sphere'' prescription for stellar collisions, assuming  there is no significant mass loss. \cite{Freitag2006b} showed that this assumption is a good approximation in old clusters with low velocity dispersion of the type considered here. Nevertheless, because this prescription is the most ``optimistic'' scenario, the results of this study are upper limits in the formation of VMSs in star clusters. 

A major source of uncertainty concerns the interior structure and radius of the collision product, particularly the effective size of the product. In cases where the VMS experiences exponential growth, the timescale between collisions is shorter than the Kelvin-Helmholtz timescale. This implies that the collision product does not have time to relax back into equilibrium before the next interaction. This is known as the ``transparency problem'' \citep{Lightman1978}. The hydrodynamics of an interaction that involves this kind of object and another star are not well understood and need to be explored in future studies. 

There have been remarkable strides in the field of modeling stellar collision products. In particular, recent studies led by \cite{Ballone2023} simulated the collision of two massive stars using the smooth particle hydrodynamics (SPH) code \texttt{Starsmasher}. The study found that the resulting stellar remnant only experienced $12$\% mass loss during the merger. \citet{Costa2022} modeled the evolution of this collision product using \texttt{PARSEC} and \texttt{MESA}, concluding that a BH in the upper-mass-gap was formed as a product of the collision. These studies represent some of the first steps towards carefully modelling collision products. They also shine light on the fact that this process is very intricate and the outcome depends on the properties of the colliding objects. Thus, detailed modeling of the hydrodynamics of stellar encounters and the properties of stellar merger products is essential to better understand VMS growth and IMBH formation from VMSs.

In future studies, we plan to research the prolonged impact and eventual collapse of these VMSs in star clusters. By running a subset of the models to a Hubble time, we aim to study cluster morphology, hypervelocity stars, and tidal tails.

\section{Acknowledgements} 
Support for E.G.P.\ was provided by the National Science Foundation Graduate Research Fellowship Program under Grant DGE-2234667.
Support for K.K.\ was provided by NASA through the NASA Hubble Fellowship grant HST-HF2-51510 awarded by the Space Telescope Science Institute, which is operated by the Association of Universities for Research in Astronomy, Inc., for NASA, under contract NAS5-26555.
Our work was supported by NASA Grant~80NSSC21K1722 and NSF Grant AST-2108624 at Northwestern University.
This research was also supported in part through the computational resources and staff contributions provided for the Quest high-performance computing facility at Northwestern University, which is jointly supported by the Office of the Provost, the Office for Research, and Northwestern University Information Technology.

\clearpage
\appendix
\section{Nested Sampling Method}

\begin{figure*}[h]
    \centering
    \includegraphics[width=0.7\textwidth]{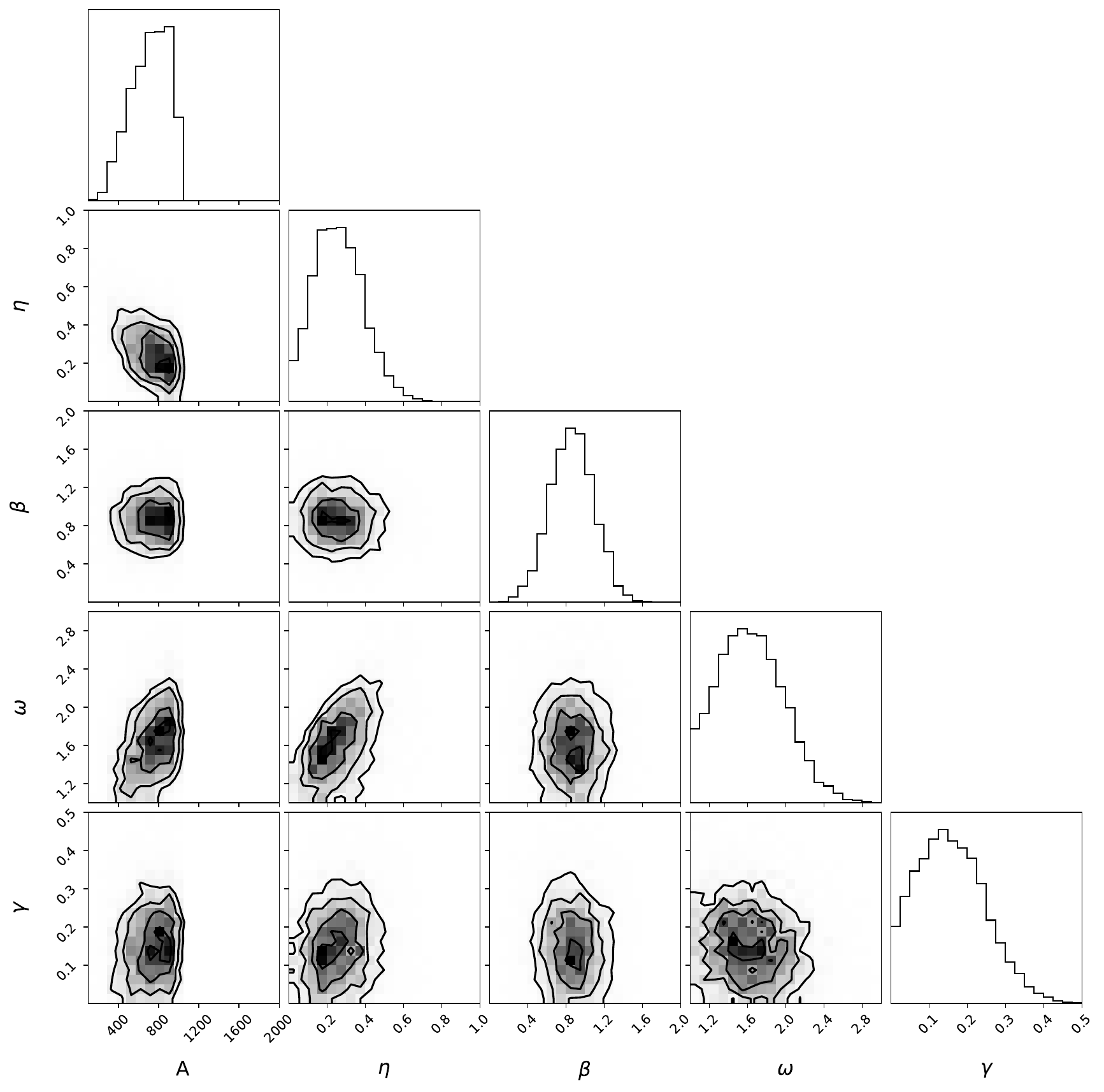}
    \caption{\label{fig:cornerplot} Corner plot for the parameters of the nested sampling fit---Equation~(\ref{eq:nested_sampling})---to the distribution of maximum stellar mass from a collisional runaway.}
\end{figure*}

As detailed in Section~\ref{sec:fitting_formulae}, we used nested sampling for the parameter estimation of Equation~(\ref{eq:nested_sampling}); for a comprehensive overview, refer to \cite{Skilling2006}. Our choice of priors includes a flat prior for $A$ ranging from $10^{-3}$--$10^3$ and from $10^{-3}$--$10^1$ for each exponent. Our sampler utilizes $10^3$ active points with a threshold of $\rm d\log(z) = 0.1$, defined as the ratio between the estimated total evidence and the current evidence. Despite experimenting with different priors, active points, and thresholds to enhance accuracy, we observe no significant differences in the results. In Figure~\ref{fig:cornerplot}, we present the corner plot derived from our parameter estimation, which shows the correlation between parameters in the off-diagonal plots and the marginal distribution of each parameter along the diagonal. From this figure we can see that the priors we provided are broad enough, suggesting that the search area is not excessively limited. To demonstrate the predictive model's performance within the parameter space explored in this paper, we reproduce Figure~\hyperref[fig:max_star]{\ref{fig:max_star}}, and overlay the mean value for each set of initial conditions using magenta, based on 1000 samples generated from the parameter distributions listed in Table~\ref{table:Table2}. As detailed in Section~\ref{sec:fitting_formulae}, the model performs well overall, with exceptions in regions characterized by high stochasticity.

\begin{figure*}
\centering
\includegraphics[width=0.7\textwidth]{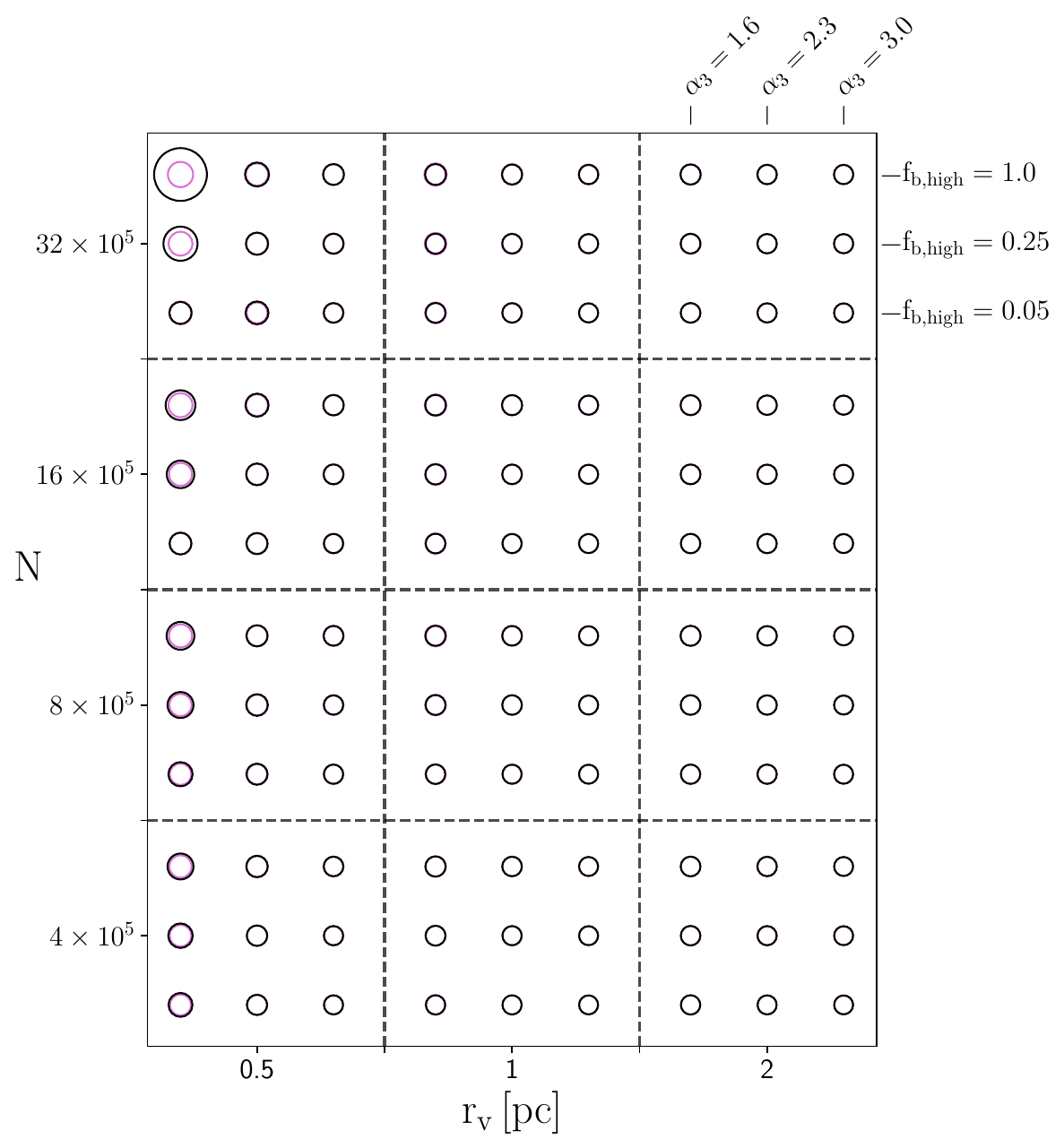}
\caption{\label{fig:model_comp} As in Figure~\hyperref[fig:max_star]{\ref{fig:max_star}}, the size of the black circles shows the average mass of the most massive VMS formed in a cluster with the given initial conditions. In this version, we also overlay in purple the predicted mean mass derived from $1000$ samples obtained from sampling values predicted by Equation~(\ref{eq:nested_sampling}). }
    
\end{figure*}

\vspace{1cm}
\bibliographystyle{aasjournal}
\bibliography{main}

\end{document}